* * *

# THE NEW WEDGE-SHAPED HUBBLE DIAGRAM OF 398 SCP SUPERNOVAE ACCORDING TO THE EXPANSION CENTER MODEL

ECM paper IX by Luciano Lorenzi

## ABSTRACT

**Following the successful dipole test on 53 SCP SNe Ia presented at SAIt2004 in Milan, this 9th contribution to the ECM series beginning in 1999 in Naples (43th SAIt meeting:"Revolutions in Astronomy") deals with the construction of the new wedge-shaped Hubble diagram obtained with 398 supernovae of the SCP Union Compilation (Kowalski et al. 2008) by applying a calculated correlation between SNe Ia absolute blue magnitude $M_B$ and central redshift $z_0$, according to the expansion center model. The ECM distance $D$ of the Hubble diagram ($cz$ versus $D$) is computed as the ratio between the luminosity distance $D_L$ and $1 + z$. Mathematically $D$ results to be a power series of the light-space $r$ run inside the expanding cosmic medium or Hubble flow; thus its expression is independent of the corresponding $z$. In addition one can have $D = D(z, h)$ from the ECM Hubble law by using the $h$ convention with an anisotropic $H_X$.**

**It is proposed to the meeting that the wedge-shape of this new Hubble diagram be confirmed independently as mainly due to the ECM dipole anisotropy of the Hubble ratio $cz/D$.**

## 1. Introduction

After the successful test of the expansion center model (Lorenzi 2004) carried out on 53 high-redshift Type Ia supernovae from the Supernova Cosmology Project (SCP: Perlmutter et al. 1999 or P99; Knop et al. 2003 or K03), here is presented the ECM construction of the new wedge-shaped Hubble diagram obtained by data from the SCP Union Compilation (Kowalski et al. 2008). In particular this large "Union" sample reports redshifts and blue magnitudes of 398 SNe Ia, or of 307 SNe Ia after selection cuts, including the distant supernovae recently observed with HST.

Let us remark that the cited papers I-II-III-IV-V-VI-VII-VIII are those of the author's references: Lorenzi 1999→2009.

## 2. Distances from the ECM equation

The **new Hubble law** (59) of paper I

$$\dot{r} = r \cdot (H + \Delta H) - R\Delta H \cos\gamma \tag{1}$$

, after substituting $H = H_0 + \Delta H, \Delta H, R$ with the formulas (37)(39) from paper I, becomes the **ECM $\dot{r}$ equation** (from eq. (3) in paper II) of the nearby Universe, expressed in Hubble units (H.u.) as follows

$$\dot{r} = H_0 \cdot r \left(\frac{1+x}{1-x}\right) \left[1 + 3q_0 \frac{(1-x)^{\frac{1}{3}}}{1+x} \cos\gamma\right] \tag{2}$$

with

$$x = \frac{3H_0 r}{c} < 1 \qquad q_0 = -\frac{H_0 R_0}{c} \qquad \cos\gamma = \sin\delta_{VC}\sin\delta + \cos\delta_{VC}\cos\delta\cos(\alpha - \alpha_{VC})$$

being

$$K_0 R_0 = a_0 = -3H_0 q_0$$

Specifically $\gamma$ is the angle between the direction ($\alpha_{VC} \approx 9^h, \delta_{VC} \approx +30^0$) of the huge void center (Bahcall & Soneira 1982), also called the expansion center or Big Bang central point (Lorenzi 1989-91-93), distant $R_0$ from the Local Group (LG) at our epoch and that $(\alpha, \delta)$ of the observed outer galaxy/group/cluster/supernova at a distance $r$ from LG, with the nearby Universe radial velocity $\dot{r}$ corrected only by the standard vector (Sandage & Tammann 1975a)(Lorenzi: paper I).

Of course $\Delta H = 0$ in eq. (1) should give the original Hubble law: $\dot{r} = H_0 r$.

Here it should be noted that the above equation (2) allows us to define at least three different cosmic distances, the following $r$, $D$ and $D_L$, which in practice have approximately the same value only for the very nearby Universe.

## 2.1 $r$ : Distance as light-space

First of all the distance $r$ in eq. (1) and (2) represents the light-space run with constant speed $c$ inside the expanding "cosmic medium" (CM hereafter) or Hubble flow. In particular such a CM flow refers to the motion of galaxies running away from the Big Bang central point, with radial velocity $\dot{R} = HR$ (cf. papers I, V and VIII). Let us rewrite the **light-space** $r$ formulation.

$$r = -c(t - t_0) \qquad with \qquad -c = \frac{\delta r}{\delta t} \tag{3}$$

In eq. (3) $t_0$ is a constant representing our epoch, which is also represented by $r = 0$; at $t_0$ the light emitted at an epoch $t$ reaches the observer at rest in the local Hubble flow, which now is more rarefied like the CM; $\delta r$ is the infinitesimal CM space covered by the light during an infinitesimal $\delta t$ of the light travel-time from the past. To all intents and purposes the source distance $r$ of eq. (3) may be considered to be equal to that of the source at the emission epoch $t$. However the cosmic medium is expanding, while light speed $c$ remains constant with respect to the local cosmic medium, as follows: $\lambda = cT \Rightarrow d\lambda = cdT \Rightarrow \lambda + \Delta\lambda = c(T + \Delta T) \Rightarrow \lambda_0 = cT_0$. In other terms the travelling light has two speeds, the former being $c$ inside CM, the latter that of the supporting expanding CM or Hubble flow. The observed velocity of this expanding CM is the derivative of $r$ to $t$, with $dt_0/dt \equiv \lambda_0/\lambda$ assumed, as shown in papers V and VIII, sections 4.7 and 2.1 respectively. That $dr/dt$ results to be $c\Delta\lambda/\lambda$, that is $\dot{r} = cz$. By introducing $z$ in eq. (2) we obtain the dimensionless **ECM** $z$ **equation** (eq. (22) of paper V or eq. (13) of paper VI or eq. (6) of paper VII)

$$z = \frac{x}{3}\left(\frac{1+x}{1-x}\right)\left[1 + 3q_0\frac{(1-x)^{\frac{1}{3}}}{1+x}\cos\gamma\right] \tag{4}$$

where we must specify

$$r \to 0 \Rightarrow x \to 0 \Rightarrow z \to 0 \Rightarrow t = t_0; H = H_0; R = R_0; K = K_0; a = a_0; q = q_0; r = D = D_L$$

$$\cos\gamma = 0 \Rightarrow z = z(x) \equiv z_0 \Rightarrow x = x(z_0) \Rightarrow r = r(z_0)$$

In fact the value $\cos\gamma = 0$ in eq. (4) leads naturally to another important convention, that is the introduction of $z_0$ to represent the **central redshift**, which must not be confused with $z(t_0) = 0$.

Moreover the previous eq. (4), with the $H_0$ and $q_0$ values obtained within the nearby Universe (paper II: $H_0 = 69.8 \pm 2.8$ $km\ s^{-1} Mpc^{-1}$; $q_0 \cong -0.0605$ from the data of Sandage & Tammann (1975)), allows a **numerical calculus of** $x$, that is of **the light-space** $r$ **as a function of the observed** $z$ **and** $\gamma$, as follows

$$x = x(z, \cos\gamma) = 3H_0 r/c \Rightarrow r = \frac{c \cdot x(z, \cos\gamma)}{3H_0} = r(z, \cos\gamma) \tag{5}$$

Note that $\cos\gamma = 0$ gives to $z_0 = 0.5$ an $x = 0.5$, that is $r \cong 716$ $Mpc$.

## 2.2 $D$ : Distance in the Hubble diagram and the $h$ convention

In 1975 Sandage & Tammann published a paper (S&T: Paper V) in which an accurate data listing of nearby galaxies (mean depth of $\sim 28$ $Mpc$) was tabled and reported in a famous wedge-shaped Hubble diagram, where the Hubble ratios appeared scattered between $\sim 30$ and $\sim 150$ $km$ $s^{-1}Mpc^{-1}$. Another wedge-shaped velocity-distance diagram, with different symbols for different methods and a covered distance depth of about 200 $Mpc$, is that of Rowan-Robinson (1988); here the Hubble constant appears to lie in the range $50 - 80$ $km\ s^{-1}Mpc^{-1}$, with a current best value in the middle of this range.

Such a wedge feature of the original Hubble diagram is well represented by eq. (2) and (4). In fact, after putting

$$D = r \cdot \left(\frac{1+x}{1-x}\right) \tag{6}$$

we can transform eq. (2) into the **ECM Hubble law**

$$cz = [H_0 - a^*(x)\cos\gamma] \cdot D = [H_0 - a_0 X(x, \cos\gamma)] \cdot D = H_X \cdot D \tag{7}$$

being

$$a^*(x) = a_0 \cdot (1-x)^{\frac{1}{3}}/(1+x) \qquad X(x, \cos\gamma) = \cos\gamma \cdot (1-x)^{\frac{1}{3}}/(1+x)$$

where both $x$ and $X$ are dimensionless variables (cf. paper V and VI); hence the above eq. (7) contains **an anisotropic angular coefficient**, that is

$$H_X = H_0(1 - \frac{a_0}{H_0} X) \tag{8}$$

As in the very nearby Universe in practice $x \to 0$, here eq. (7) gives $a^* \simeq a_0$, that is $H_X(\gamma = 0^0) \cong 57$ $km\ s^{-1}Mpc^{-1}$ and $H_X(\gamma = 180^0) \cong 83$ $km\ s^{-1}Mpc^{-1}$ with $a_0 \cong 12.7$ $km\ s^{-1}Mpc^{-1}$(paper V: section 4.6).

The MacLaurin Series applied to (6) and the ECM Hubble law (7) give $D$ both in terms of a power series of the light-space $r$,

$$D = r + 2\frac{3H_0}{c}r^2 + 2\frac{9H_0^2}{c^2}r^3 + ... \tag{9}$$

and as a function of $z_0$, that is the ratio between the central velocity $cz_0$ and the constant $H_0$:

$$D = \frac{cz_0}{H_0} = D(z_0) \tag{10}$$

The eqs. (6)(9)(10) represent the **distance** $D$ of the wedge-shaped Hubble diagram of eq. (7).

At the same time **the ECM Hubble law (7) is able to substantiate the powerful** $h$ **convention** (Zeilik & Smith 1988) for large-scale surveys of radial distance $D$ in H.u., by using a variable $h = h(X)$ tied to the ECM apparent anisotropy (cf. paper II, secion 1.2) and the correct $z$, obtained after subtracting from the observed heliocentric redshift the kinematic component due to the entire motion of the Sun with respect to the Hubble flow traced by the CMB (Lorenzi 1993, 1999a, 2008, 2009). So we confirm the following useful formula:

$$D = \frac{cz}{100\ km\ s^{-1}Mpc^{-1}}h^{-1} \qquad with \qquad h = \frac{H_X}{100\ km\ s^{-1}Mpc^{-1}} \tag{11}$$

## 2.3 $D_L$ : Luminosity distance and correlated absolute magnitude $M$

Papers V and VI have empirically confirmed the ECM even for Abell clusters of Richness 3 and Type Ia supernovae from SCP, up to a light-space distance $r$ of $\sim 1000\ Mpc$. Here the luminosity distance $D_L$ has been successfully represented by the following **ECM** $D_C$ **multiple formula**

$$D_C = D(1+z) = \frac{xc}{3H_0}\left(\frac{1+x}{1-x}\right)(1+z) = \frac{cz(1+z)}{H_0}(1 - \frac{a_0}{H_0}X)^{-1} \tag{12}$$

Consequently, with $D_C \equiv D_L$ assumed, the distance $D$ of the Hubble diagram can be simply inferred from the position

$$D = \frac{D_L}{1+z} \tag{13}$$

when one knows the absolute magnitude $M$, that is

$$M = m - 5\log D_L - 25 \tag{14}$$

By combining the canonic eq. (14) in H.u. with (12) and (13), we can obtain the **ECM** $M$ **equation,** written in a double form:

$$M = m - 5\log\left[\frac{xc}{3H_0}\left(\frac{1+x}{1-x}\right)(1+z)\right] - 25 \tag{15}$$

$$M = m - 5\log\left[cz(1+z)\right] + 5\log H_0 + \Delta - 25 \tag{16}$$

In (14)(15)(16) $m$ and $z$ are the observed magnitude and redshift within the Hubble flow; in (15) $x = x(z, \cos\gamma)$ from eq. (4); $\Delta$ of eq. (16) results to be a power series of $X(x, \cos\gamma)$, as follows

$$\Delta = 5\log(1 - \frac{a_0}{H_0}X) = -\frac{5a_0}{H_0}\log e \cdot (X + \frac{a_0}{H_0}\frac{X^2}{2} + \frac{a_0^2}{H_0^2}\frac{X^3}{3} + .....) \tag{17}$$

Eq. (16) can be simplified by introducing the central redshift $z_0$ corresponding to $z$ of eq. (4) with $\cos\gamma = 0$, that is $H_X = H_0$. In this case, being

$$X \equiv 0 \Rightarrow \Delta = 0 \Rightarrow z = z_0 \Rightarrow m = m_0(z_0) \Rightarrow D_L = D_L(m_0) \tag{18}$$

, we also obtain the **ECM** $M(z_0)$ **equation**, in the form

$$M = m_0 - 5\log\left[cz_0(1+z_0)\right] + 5\log H_0 - 25 \tag{19}$$

## 3. Construction of the ECM Hubble diagram of 398 SCP supernovae

The main aim of the present work is the application of the above formulae to the largest available sample of homogeneous datasets. The SCP "Union" SNe Ia compilation holds such a sample, bringing together data from 414 SNe (Kowalski et al 2008: Table 11) drawn from 13 independent datasets, of which 398 SNe have both the required redshifts $z$ and blue magnitudes $m_B^{\max}$ listed, while a wide subsample of 307 SNe Ia pass usability cuts. Note that here the redshifts $z$ are referred to the CMB; hence they include the correction due to the standard motion of the Local Group, without taking into account the ECM 3K dipole able to generate a fictitious vector $v_f$ (Lorenzi 1993, 1999a, 2008). As the involved correction to $z$ is about 0.001 on average, the $z$ of the distant supernovae in effect do not suffer an imprecise correction; it is different for the very nearby SNe, whose redshifts in the Hubble diagram should be corrected only for the Sun's velocity inside the Local Group (by the standard vector of S&T (1975)), because our LG belongs to a large local cosmic flow also running almost in the same direction (cf. p. 19 of paper I).

On the whole the present analysis aims directly to construct the ECM Hubble diagram, of course without using $\cos\gamma$, but showing in any case that the diagram's wedge-shape is due to the ECM dipole anisotropy. A further and crucial confirmation of the model is expected by introducing the supernova astronomical coordinates, that is to say $\cos\gamma$, into the ECM analysis both for SCP Union (Kowalski et al. 2008) and the SCP "Union2" shown at 2010 AAS (Rubin et al. 2010).

### 3.1 Search for a correlation between $M_B$ and central redshift $z_0$

Initially the ECM Hubble law (7) was tested over the 398 SCP supernovae (Kowalski et al. 2008: Table 11), by assuming $H_0$ as the average Hubble ratio, that is

$$H_0 = \langle H_X \rangle = \langle \frac{cz}{D} \rangle \tag{20}$$

The procedure, based on the mean eq. (20) in H.u. with $D$ derived from (13) and $D_L = 10^{0.2(m_B^{\max} - M_B) - 5}$, was applied to five large $z$ bins of the Hubble flow, precautionally excluding the nearby SNe with $z < 0.05$; hence, once $M_B$ or a resulting $H_0 \cong 70$ H.u. are fixed, the value of $H_0$ or $M_B$ follow. Conditions and results of that first check are listed below, in Table 1.

**Table 1**

| $z$ bins | N | $\langle z \rangle$ | $H_0 = H_0(M_B = -19.5)$ | $M_B = M_B(H_0 \cong 70)$ |
|---|---|---|---|---|
| $0.05 \leq z \leq 0.5$ | 145 | 0.334 | **62** | **−19.25** |
| $0.25 \leq z \leq 0.75$ | 197 | 0.492 | **67** | **−19.41** |
| $0.05 \leq z \leq 1.552$ | 308 | 0.568 | **70** | **−19.49** |
| $0.5 \leq z \leq 1.0$ | 142 | 0.706 | **74** | **−19.63** |
| $0.75 \leq z \leq 1.25$ | 67 | 0.916 | **81** | **−19.81** |

The strong variation of the $H_0$ value in the $4^{th}$ column, corresponding to the assumed $M_B = -19.5$ (cf. paper VI), clearly rules out the possibility of a constant value of the SNe Ia absolute blue magnitude $M_B$. On the other hand the constant value of $H_0$ gives to SNe Ia a variable intrinsic luminosity, which clearly increases with depth or central redshift, according to the ECM.

Owing to the clear result in Table 1 and in order to construct a correlation between $M_B$ and the central redshift $z_0$ according to (19), the same "$z$ bins" procedure has been applied to a **normal ECM $M$ equation**, that is eq. (16) with $H_0 = 70$ H.u. and $\langle \Delta \rangle = 0$ assumed, as follows

$$\langle M_B \rangle = \langle m_B^{\max} \rangle - 5 \langle \log [cz(1+z)] \rangle + 5 \log H_0 - 25 \tag{21}$$

In this case the check is more useful than the previous one, first of all because eq. (21) gives directly $\langle M_B \rangle$ with its standard deviation; furthermore the ECM eq. (21) seems to be statistically powerful, if the $z$ scattering due to unsuitable corrections of Sun kinematics in the CMB is assumed to be neutralized like $\Delta$ by the normal point, apart from any anisotropies of SNe Ia distribution in the sky plus a $H_0$ imprecision of about $\pm 0.1$ magnitudes. In other words eq. (21) seems able to produce $M_B$ values correlated to the central redshift $z_0$ of eq. (19). Thus all the available SCP

SNe Ia of the Union 2008, 91 nearby SNe with $z \leq 0.05$ included, have been taken into account. Table 2 in the Appendix lists the results of the mean; it reports 30 normal points, including all the 398 SNe listed in Table 11 of the 2008 SCP paper (Kowalski et al. 2008) and corresponding to 30 "$z$ bins". In particular the first 5 columns of Table 2 hold numerical values derived from the observed $z$ and $m_B^{\max}$ listed within the above SCP Union 2008; the values referring to each $z$ bin are in the order: $z$ range, number N of the SNe included in the normal point; unweighed mathematical mean $\langle m_B^{\max} \rangle$ of the observed SN blue magnitudes $m_B^{\max}$; absolute magnitude $\langle M_B \rangle$ resulting from the normal ECM $M$ equation (21) applied to the bin, with $H_0 = 70$ H.u. assumed; standard deviation $s$ of the least square fitting carried out on the bin. The $6^{th}$ column of Table 2 reports the mathematical mean $\langle z \rangle$ of the observed redshifts of the $z$ bin, while the last columns, $7^{th}$, $8^{th}$, $9^{th}$, include three different distance values, corresponding to an assumed central redshift $z_0 \equiv \langle z \rangle$ with $H_0 = 70$ H.u.. These are in the order: value of the dimensionless variable $x = x(z_0)$, inferred as in (5) from the ECM $z$ equation (4); value in $Mpc$ of the light-space distance $r = r(z_0)$ connected to the $x$ value by $x = 3H_0 r/c$ according to procedure (5) applied to eq. (2) or (4); value in $Mpc$ of the distance $D$ of the wedge-shaped Hubble diagram, obtained with eq. (6) or (10), that is as the function $D = D(z_0)$ of the central redshift through $x(z_0)$.

The resulting normal points, plotted in Figure 1 as $\langle M_B \rangle$ versus $\langle z \rangle$, clearly point to a fitted trend line, whose equation formally should give for any $z$ its $M_B$ as a function of the central redshift $z_0$, or the corresponding distance $D = cz_0/H_0$ as in Figure 2, if $z_0$ is assumed to refer to the line fitting the $\langle z \rangle$ points. The line equation below,

$$M_B(z_0) = A_0 + A_1 z_0 + A_2 z_0^2 = d_0 + d_1 D + d_2 D^2 = M_B(D) \tag{22}$$

, with $A_0 \cong -18.77$; $A_1 \cong -1.421$; $A_2 \cong +0.3589$ and $d_0 = A_0$; $d_1 = A_1 H_0/c$; $d_2 = A_2 H_0^2/c^2$, follows from the automatic fitting.

In the same way, according to procedure (5) applied to eq. (2) or (4) with $\cos\gamma = 0$, that is with $z = z_0$, an alternative plot of normal points $\langle M_B \rangle$ versus $x = x(z_0)$ or $r = r(z_0)$ can be constructed; it appears in Figure 3 and Figure 4, where the fitted trend line appears better represented by a third degree equation, that is

$$M_B(x) = B_0 + B_1 x + B_2 x^2 + B_3 x^3 = C_0 + C_1 r + C_2 r^2 + C_3 r^3 = M_B(r) \tag{23}$$

, with $B_0 \cong -18.78$; $B_1 \cong -0.4523; B_2 \cong -0.5338$; $B_3 \cong -2.006$ and $C_0 = B_0$; $C_1 = 3B_1 H_0/c$; $C_2 = 9B_2 H_0^2/c^2$; $C_3 = 27 B_3 H_0^3/c^3$, again obtained from the automatic fitting. Here the curve agrees with that found in paper VI for 33 SNe Ia of K03 (cf. paper VI-integral version: Fig. 5), however with a systematic shift of about 0.3 magnitudes limited to the nearby Universe.

### 3.2 Construction of the Hubble diagram

The above equation (22), that expresses $M_B(D)$, has a crucial role in the construction of the SNe Ia Hubble diagram, which requires the distance $D$ to combine with the observed redshift as $cz$. In fact it is now possible to extract numerically just the distance $D$ from the canonic eq. (14), that becomes the following:

$$d_2 D^2 + d_1 D + d_0 = m_B^{\max} - 5 \log \left[ D(1+z) \right] - 25 \tag{24}$$

The numerical solution point by point of eq. (24), here applied to 398 SNe Ia with $z$ and $m_B^{\max}$ listed in Table 11 of the SCP Union Compilation (Kowalski et al. 2008), gives the value of $D$. Once found, one can infer numerically also the corresponding values of the distance indicator $x$ and the light-space distance $r$.

Finally, two resulting wedge-shaped Hubble diagrams in H.u. are obtained by plotting $cz$ versus $D$ for the 398 SNe, in Figure 5, and the 307 SNe passing usability cuts, in Figure 6. Here we look at the Hubble diagram of the Deep Universe. Table 3abcdefghi in the Appendix lists the values in H.u. of $D$ ($3^{rd}$column) and $cz$ ($2^{nd}$column) of 249 SNe Ia ($1^{st}$column: Name), lying in the distance range 800 $Mpc < D <$ 8000 $Mpc$, from the 307 SNe selected by the SCP Union.

## 4. ECM analysis of the wedge-shaped Hubble diagram

The diagrams in Fig. 5 and Fig. 6 have a wedge shape, whose amplitude with increasing depth is very large, indeed. In order to verify the accordance with the model, it is necessary to compare the observed wedge shape to the calculated one. In practice we should carry out an (O-C) procedure. That has been done by calculating the wedge amplitude foreseen by the ECM Hubble law (7) through a numerical **simulation** of the maximum scattering of $cz$ around the central value $cz_0$. To this end, let us analyse the observed Hubble flow or CM as follows:

$$z_0 = \frac{\lambda_0 - \lambda_e}{\lambda_e} = \frac{T_e}{T_0} - 1 \tag{25}$$

$$cz = cz_0 + c\Delta z_0 \tag{26}$$

$$\Delta z_0 = -\frac{T_e}{T_0} \left( \frac{\Delta T_0}{T_0 + \Delta T_0} \right) \tag{27}$$

As the **ECM** gives

$$cz_0 = H_0 D \qquad c\Delta z_0 = -a_0 D X \tag{28}$$

, after fixing $x$, then $r$ and $D$, solely the dimensionless $X$ varies owing to the variation of $\cos\gamma$ between 1 and $-1$. In this case:

$$-(1-x)^{\frac{1}{3}}/(1+x) \le X \le (1-x)^{\frac{1}{3}}/(1+x) \tag{29}$$

$$c\Delta z = c\Delta(\Delta z_0) = -a_0 D \Delta X \tag{30}$$

$$r = \frac{cx}{3H_0} \qquad cz_0 = \frac{cx}{3}\left(\frac{1+x}{1-x}\right) \qquad D = \frac{cx}{3H_0}\left(\frac{1+x}{1-x}\right) \tag{31}$$

$$(\Delta X)^{\max} = 2(1-x)^{\frac{1}{3}}/(1+x) \tag{32}$$

$$c\,|\Delta z|^{\max} = a_0 D (\Delta X)^{\max} \tag{33}$$

Now we add to the above **formulae** the following ones, which practically, being based solely on the eqs. (25)(26)(27), are **unaffected by the ECM** (cf. section 6 of paper VII).

$$\Delta T_0 = \Delta T_D \cos\gamma \tag{34}$$

$$\frac{\Delta z_0}{1+z_0} = -\frac{\Delta T_0}{T_0 + \Delta T_0} \tag{35}$$

$$\gamma = 0 \Rightarrow \frac{\Delta z_0}{1+z} = -\frac{\Delta T_0}{T_0} \equiv \frac{v_f}{c} \Rightarrow v_f = \frac{c\Delta z_0}{1+z} \tag{36}$$

The last equation of (36) gives the value of the fictitious velocity $v_f$ observed towards the expansion center. Its value in H.u., obtained **within the ECM**, is listed in the $7^{th}$ column of Table 4, while the previous 5 columns present the other simulated values in H.u. of $r, cz_0, D, (\Delta X)^{\max}, c\,|\Delta z|^{\max}$ from the above eqs. (31)(32)(33) corresponding to the $x$ value of the first column. The last row of Table 4, where $x = 0.999999225$, reports the **ECM values extrapolated to the CMB**, whose fictitious velocity results to be only of the order $-250$ $km\ s^{-1}$. Curiously this value, that is in accordance with the observed 3K anisotropy and a local cosmic flow of about 530 $km\ s^{-1}$ (cf. Lorenzi 1993, 1999a, 2008), is the same of the nearby Universe at $D \cong 20$ $Mpc$, while at $D \approx 10000$ $Mpc$ the corresponding value of $v_f$ reaches a maximum of about $-13000$ $km\ s^{-1}$.

On the whole, the ECM simulation is able to reproduce the variable wedge-shape of the Hubble diagram at different depths, as summarized in the table below.

**Table 4**

| $x$ | $r$ | $cz_0$ | $D$ | $(\Delta X)^{\max}$ | $c\,\lvert\Delta z\rvert^{\max}$ | $v_f$ |
|---|---|---|---|---|---|---|
| 0.00070 | 0.9993 | 70 | 1 | 1.998 | 25 | $-13$ |
| 0.00691 | 9.865 | 700 | 10 | 1.982 | 252 | $-126$ |
| 0.013633 | 19.46 | 1400 | 20 | 1.964 | 499 | $-248$ |
| 0.026569 | 37.93 | 2800 | 40 | 1.931 | 981 | $-487$ |
| 0.038883 | 55.51 | 4200 | 60 | 1.900 | 1448 | $-716$ |
| 0.050637 | 72.29 | 5600 | 80 | 1.871 | 1901 | $-936$ |
| 0.061884 | 88.34 | 7000 | 100 | 1.844 | 2342 | $-1148$ |
| 0.342820 | 489.4 | 70000 | 1000 | 1.295 | 16445 | $-6818$ |
| 0.485378 | 692.9 | 140000 | 2000 | 1.079 | 27407 | $-9642$ |
| 0.631939 | 902.1 | 280000 | 4000 | 0.8783 | 44617 | $-11997$ |
| 0.710716 | 1015 | 420000 | 6000 | 0.7732 | 58918 | $-12793$ |
| 0.760902 | 1086 | 560000 | 8000 | 0.7049 | 71622 | $-13029$ |
| 0.7959346 | 1136 | 700000 | 10000 | 0.6556 | 83266 | $-13026$ |
| 0.9 | 1285 | 1708817 | 24412 | 0.4886 | 151476 | $-11747$ |
| 0.99 | 1413 | 19687373 | 281248 | 0.2165 | 773401 | $-5915$ |
| 0.999999225 | 1427.582 | $\simeq 2.6 \times 10^{11}$ | $\simeq 3.7 \times 10^{9}$ | 0.0092 | $\simeq 4.3 \times 10^{8}$ | $-250$ |

Finally, Figure 7 and Figure 8 present two plots of the results in Table 4: the former refers to the nearby cosmic region with $D \leq 80\ Mpc$ and the latter to the Deep Universe with $D \leq 8000\ Mpc$. We obtain two simulated ECM Hubble diagrams in H.u., where the plotted points corresponding to each tabled $D$ are the central $cz_0$, the upper $cz_0 + \frac{c|\Delta z|^{\max}}{2}$ and the lower $cz_0 - \frac{c|\Delta z|^{\max}}{2}$ . From their comparison, it results clearly that the wedge amplitude has to decrease with depth, until $H_X \to H_0$ for $x \to 1$.

## 5. Conclusions

This paper validates the wedge-shaped Hubble diagram predicted by the espansion center model. In fact the diagrams of the Deep Universe in Fig. 5 and Fig. 6 are in good accordance with that simulated in Fig. 8, as the observed amplification is certainly due to various sources of background noise. Therefore this $9^{th}$ ECM contribution, based above all on a large sample of SCP data obtained in space with HST, is further confirmation of the cosmic expansion center, following the ground-based astronomical proof collected in about half a century.

At this point one cannot desist from pressing the international astronomical community to pronounce itself once and for all on the subject. In conclusion the author extends an invitation to all astronomers to analyse independently the presented new wedge-shaped Hubble diagram, in order to confirm, or confute, the dipole anisotropy of the Hubble ratio $cz/D$ at any cosmic depth.

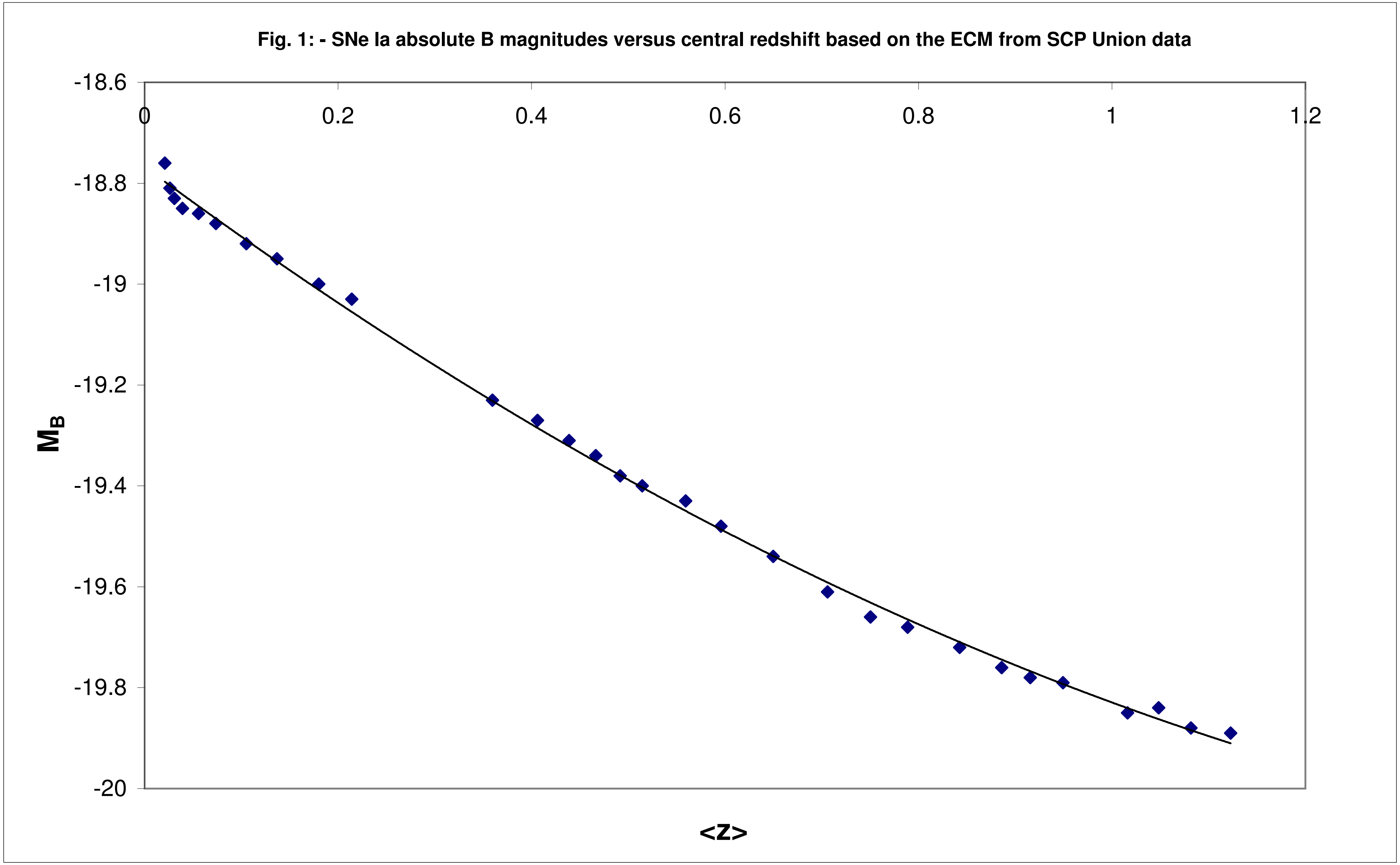

Fig. 1: - SNe Ia absolute B magnitudes versus central redshift based on the ECM from SCP Union data
-18.6
-18.8
-19
-19.2
-19.4
-19.6
-19.8
-20
0
0.2
0.4
0.6
0.8
1
1.2
$M_B$
<z>

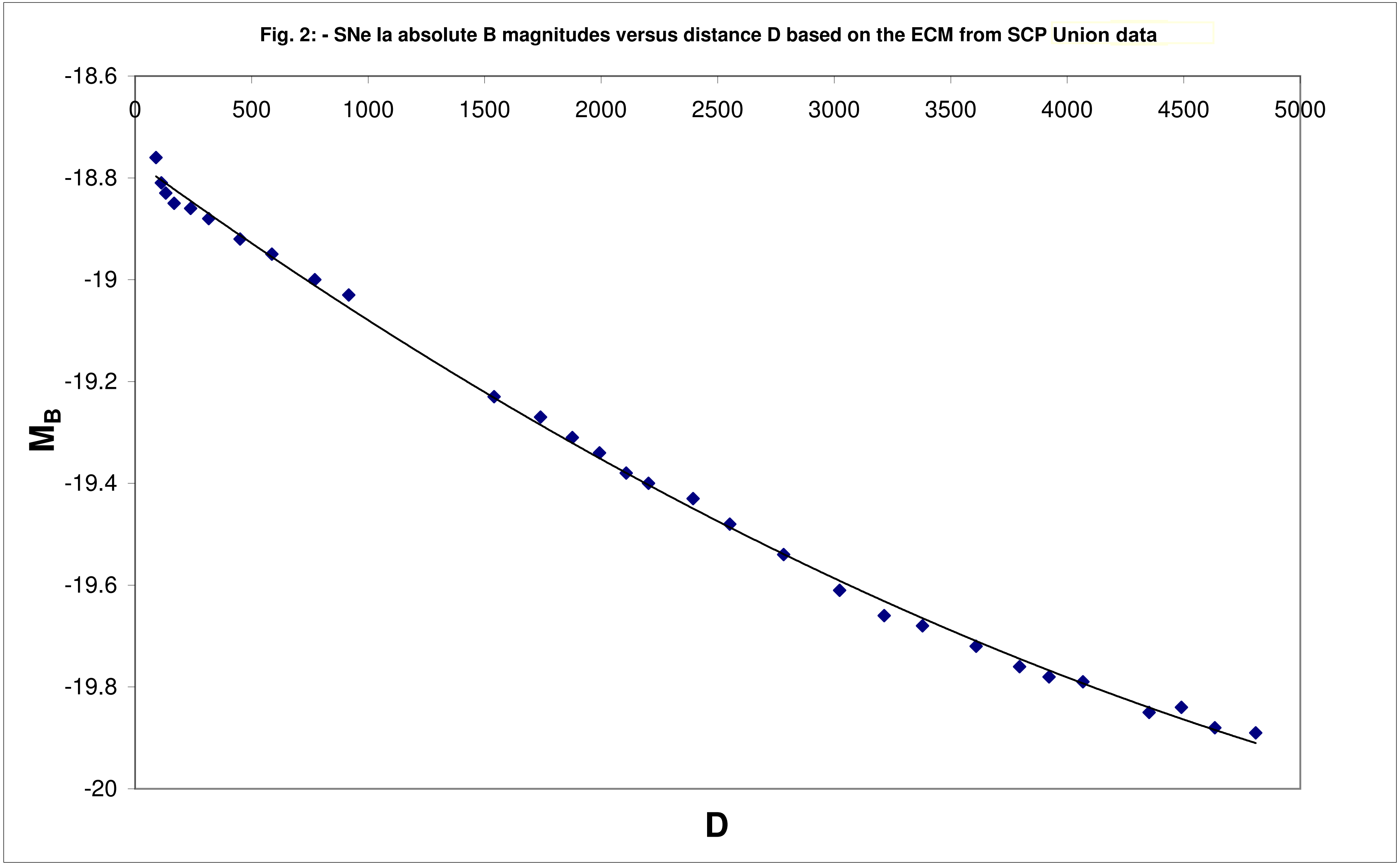
Fig. 2: - SNe Ia absolute B magnitudes versus distance D based on the ECM from SCP Union data
-18.6
-18.8
-19
-19.2
-19.4
-19.6
-19.8
-20
0
500
1000
1500
2000
2500
3000
3500
4000
4500
5000
$M_B$
D

Fig. 3: - SNe Ia absolute B magnitudes versus the distance indicator x based on the ECM from SCP Union data

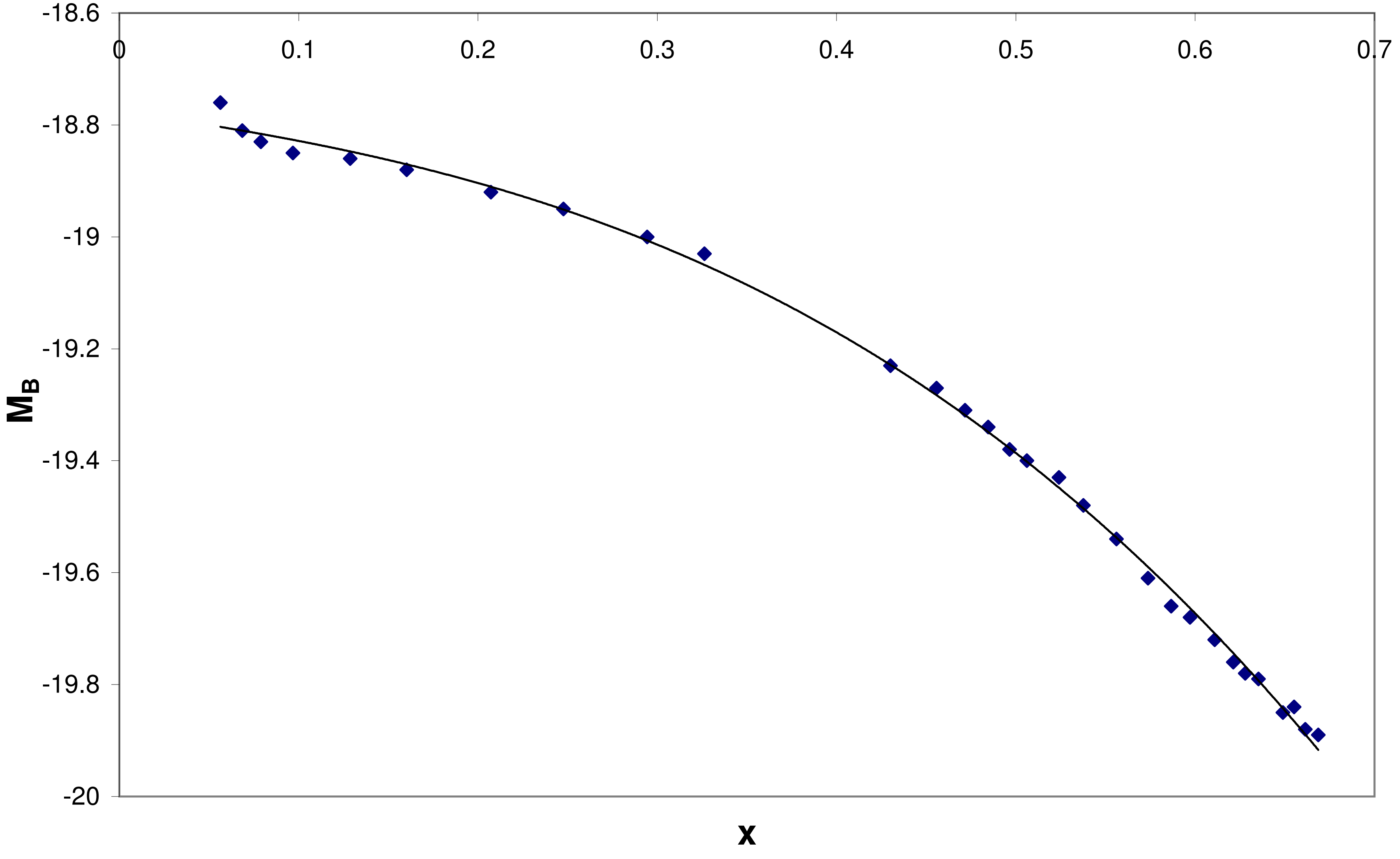

**Fig. 4: - SNe Ia absolute B magnitudes versus light-space distance r based on the ECM from SCP Union data**

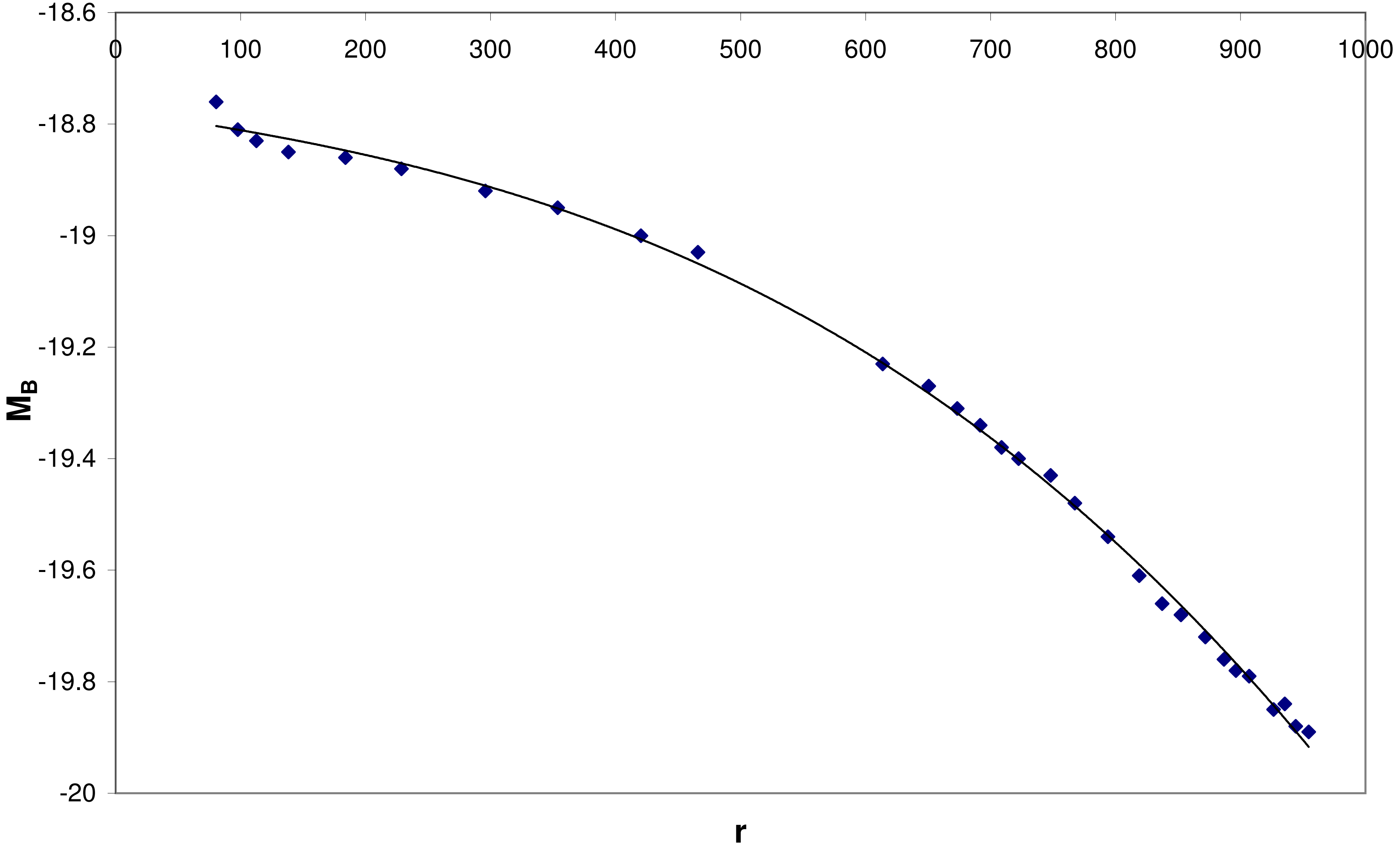

Fig. 5: - New Hubble diagram of 398 SCP supernovae

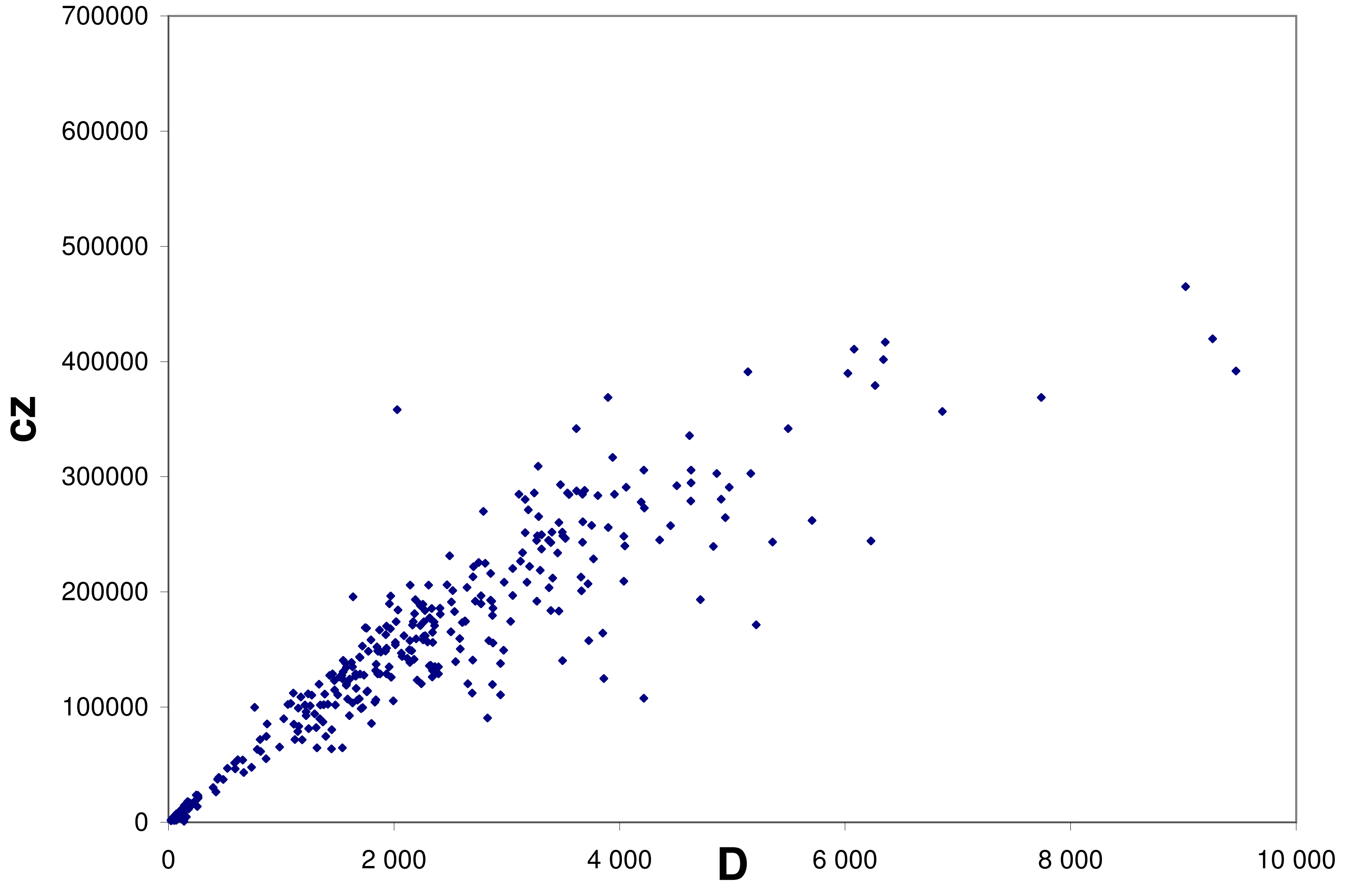

**Fig. 6: - New Hubble diagram of 307 SCP supernovae**

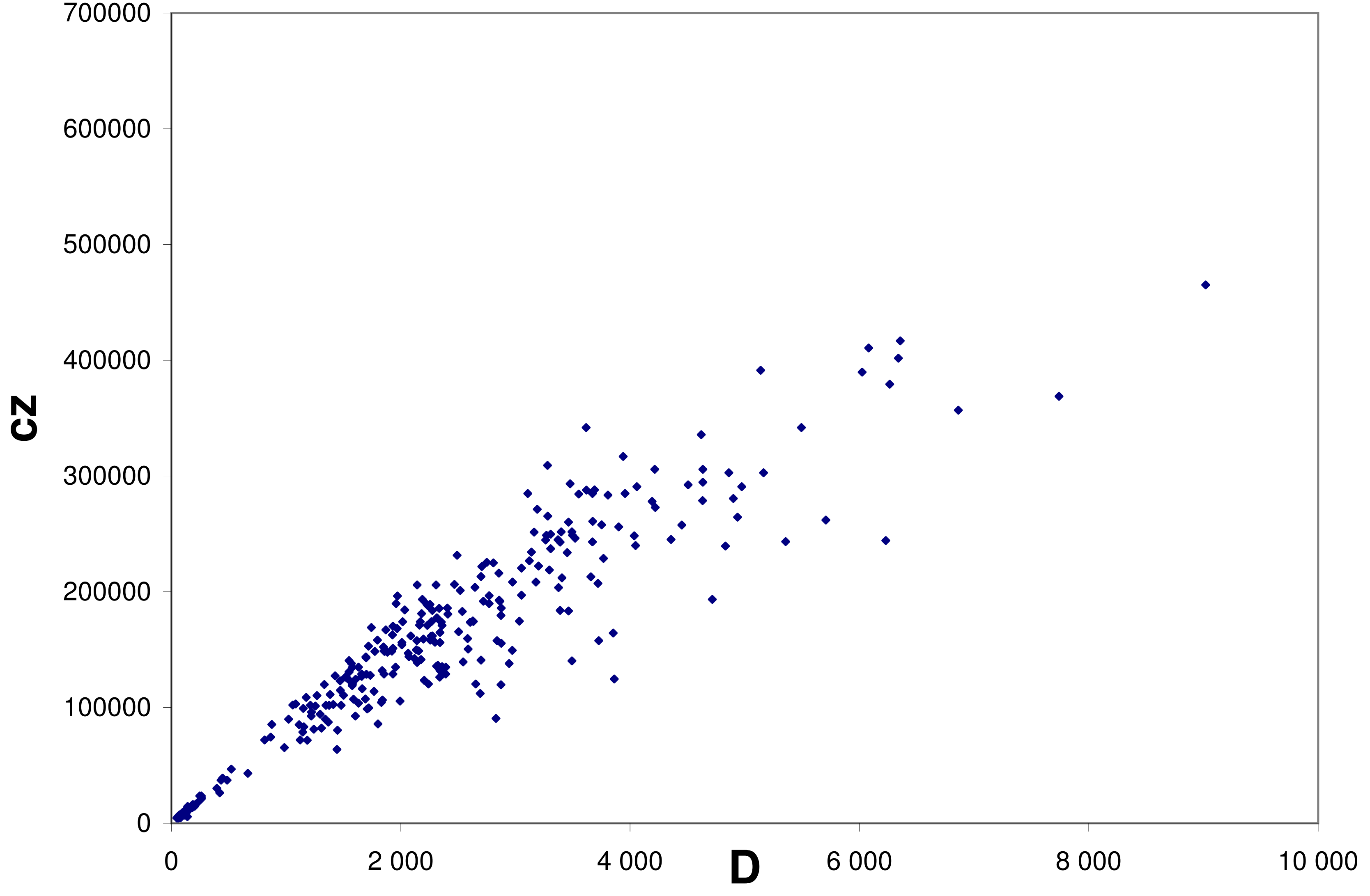

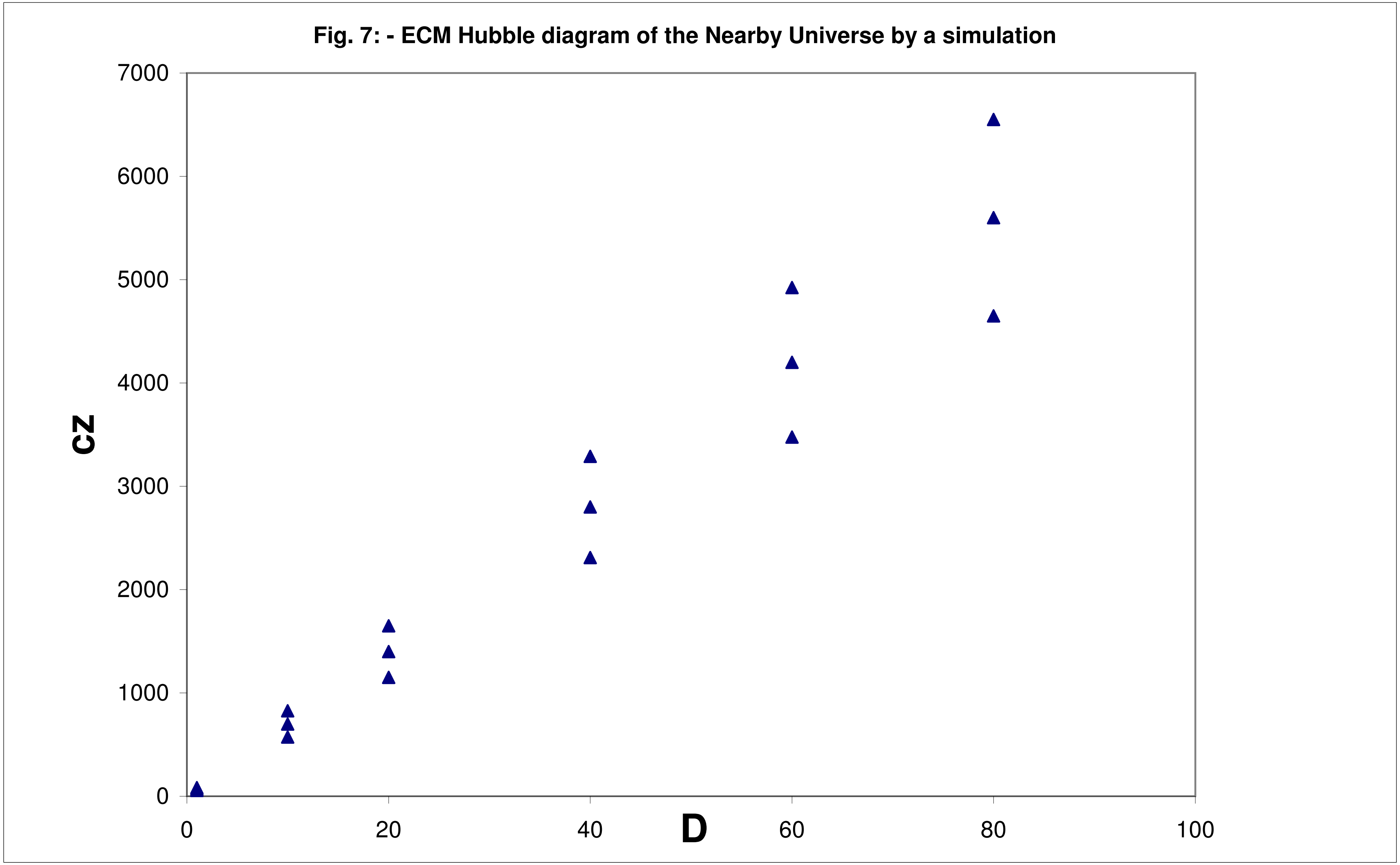
Fig. 7: - ECM Hubble diagram of the Nearby Universe by a simulation
7000
6000
5000
4000
3000
2000
1000
0
cz
0
20
40
D
60
80
100

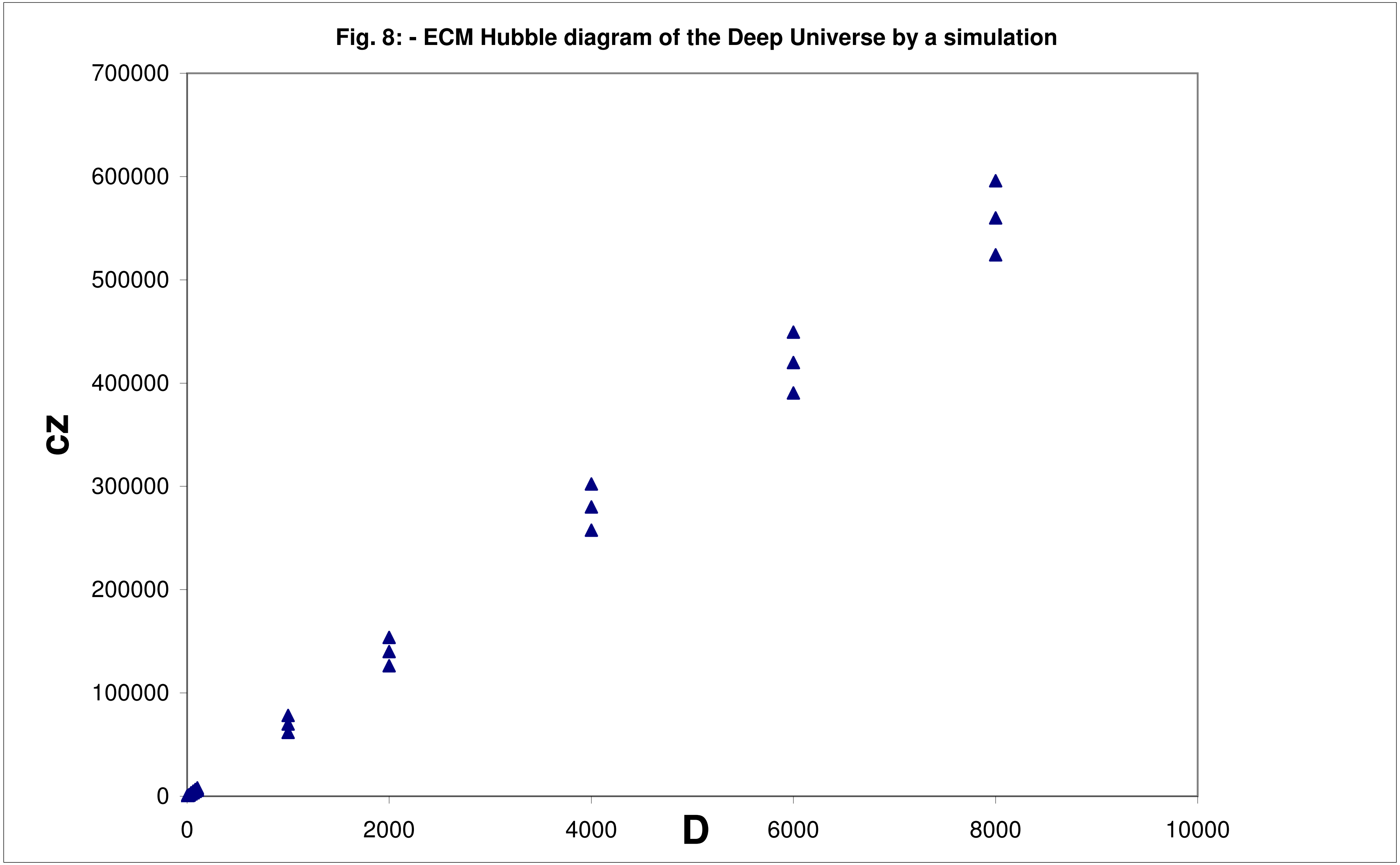
Fig. 8: - ECM Hubble diagram of the Deep Universe by a simulation
cz
700000
600000
500000
400000
300000
200000
100000
0
0
2000
4000
D
6000
8000
10000

**Acknowledgements**

This work was made possible thanks to the SCP Union Compilation. The author would like to thank all the members of the SCP team, in particular for making the SNe data available on line in "arXiv:0804.4142v1 [astro-ph] 25 Apr 2008".

Special acknowledgements are reserved for the Local Organizing Committee of SAIt2010 at Capodimonte Astronomical Observatory in Naples, both for the successful meeting and all the kind attention and support given to the present contribution.

A final word of gratitude goes to a dear friend Francesco Chiapello for all his encouragement.

REFERENCES

Bahcall, N.A. and Soneira, R.M. 1982, ApJ 262, 419

Knop, R.A. et al. 2003, ApJ 598, 102 (K03)

Kowalski, M. et al. 2008, arXiv:0804.4142v1 [astro-ph] 25 Apr 2008→ApJ 686, 749

Lorenzi, L. 1989, Contributo N. 0, CSA-Mondovì, Italy (unpublished)

1991, Contributo N. 1, CSA-Mondovì, Italy

1993, in 1995 MemSAIt, 66, 249

1999a, arXiv:astro-ph/9906290v1 17 Jun 1999,

in 2000 MemSAIt, 71, 1163 (paper I: reprinted in 2003, MemSAIt, 74)

1999b, arXiv:astro-ph/9906292v1 17 Jun 1999,

in 2000 MemSAIt, 71, 1183 (paper II: reprinted in 2003, MemSAIt, 74)

2002, in 2003 MemSAIt, 74, 480 (paper III-partial version),

http://sait.oat.ts.astro.it/MSAIt740203/PDF/poster/39_lorenzil_01_long.pdf

(paper III-integral version)

2003a, MemSAIt Suppl. 3, 277 (paper IV)

2003b, MemSAIt Suppl. 3,

http://sait.oat.ts.astro.it/MSAIS/3/POST/Lorenzi_poster.pdf (paper V)

2004, MemSAIt Suppl. 5, 347 (paper VI)

2008, www.sait.it, Archivio Eventi, 2008-LII Congresso Nazionale della SAIt,

http://terri1.oa-teramo.inaf.it/sait08/slides/I/ecmcm9b.pdf (paper VII)

2009, www.sait.it, Archivio Eventi, 2009-LIII Congresso Nazionale della SAIt,

http://astro.df.unipi.it/sait09/presentazioni/AulaMagna/08AM/lorenzi.pdf

(paperVIII)

Perlmutter, S., et al. 1999, ApJ 517, 565 (P99)

Rowan-Robinson, M. 1988, Space Science Review 48, 1

Rubin, D. et al. 2010, SCP "Union 2" Shown at 2010 AAS

Sandage, A., Tammann G.A. 1975, ApJ 196, 313 (S&T: Paper V)

Zeilik, M., Smith, E.v.P. 1987, Introductory Astronomy and Astrophysics,

CBS College Publishing

## APPENDIX

**Table 2**

| $z$ range | N | $\langle m_B^{\max} \rangle$ | $\langle M_B \rangle$ | $s$ | $\langle z \rangle$ | $x$ | $r$ | $D$ |
|---|---|---|---|---|---|---|---|---|
| $z \leq 0.05$ | 91 | 15.69 | $-18.76 \pm 0.09$ | 0.848 | 0.0210 | 0.0563 | 80.4 | 90.0 |
| $z \leq 0.10$ | 103 | 15.97 | $-18.81 \pm 0.09$ | 0.817 | 0.0262 | 0.0685 | 97.8 | 112.2 |
| $z \leq 0.15$ | 108 | 16.15 | $-18.83 \pm 0.08$ | 0.802 | 0.0308 | 0.0789 | 112.6 | 131.9 |
| $z \leq 0.20$ | 115 | 16.41 | $-18.85 \pm 0.08$ | 0.785 | 0.0392 | 0.0968 | 138.2 | 167.8 |
| $z \leq 0.25$ | 126 | 16.85 | $-18.86 \pm 0.07$ | 0.763 | 0.0556 | 0.1287 | 183.8 | 238.1 |
| $z \leq 0.30$ | 137 | 17.25 | $-18.88 \pm 0.07$ | 0.741 | 0.0738 | 0.1602 | 228.7 | 316.0 |
| $z \leq 0.35$ | 156 | 17.85 | $-18.92 \pm 0.06$ | 0.727 | 0.1052 | 0.2072 | 295.8 | 450.4 |
| $z \leq 0.40$ | 177 | 18.41 | $-18.95 \pm 0.06$ | 0.712 | 0.1369 | 0.2477 | 353.6 | 586.5 |
| $z \leq 0.45$ | 208 | 19.06 | $-19.00 \pm 0.05$ | 0.684 | 0.1800 | 0.2944 | 420.3 | 771.0 |
| $z \leq 0.50$ | 235 | 19.52 | $-19.03 \pm 0.05$ | 0.662 | 0.2141 | 0.3263 | 465.8 | 917.1 |
| $0.05 \leq z \leq 0.55$ | 167 | 22.10 | $-19.23 \pm 0.04$ | 0.420 | 0.3595 | 0.4300 | 613.8 | 1540 |
| $0.10 \leq z \leq 0.60$ | 174 | 22.54 | $-19.27 \pm 0.04$ | 0.436 | 0.4062 | 0.4557 | 650.5 | 1740 |
| $0.15 \leq z \leq 0.65$ | 191 | 22.74 | $-19.31 \pm 0.04$ | 0.438 | 0.4380 | 0.4717 | 673.4 | 1876 |
| $0.20 \leq z \leq 0.70$ | 199 | 22.90 | $-19.34 \pm 0.04$ | 0.447 | 0.4663 | 0.4846 | 691.8 | 1993 |
| $0.25 \leq z \leq 0.75$ | 197 | 23.03 | $-19.38 \pm 0.03$ | 0.426 | 0.4919 | 0.4965 | 708.8 | 2107 |
| $0.30 \leq z \leq 0.80$ | 197 | 23.14 | $-19.40 \pm 0.03$ | 0.420 | 0.5144 | 0.5061 | 722.5 | 2203 |
| $0.35 \leq z \leq 0.85$ | 191 | 23.35 | $-19.43 \pm 0.03$ | 0.405 | 0.5592 | 0.5240 | 748.1 | 2394 |
| $0.40 \leq z \leq 0.90$ | 180 | 23.50 | $-19.48 \pm 0.03$ | 0.362 | 0.5958 | 0.5376 | 767.5 | 2552 |
| $0.45 \leq z \leq 0.95$ | 162 | 23.71 | $-19.54 \pm 0.03$ | 0.352 | 0.6498 | 0.5561 | 793.9 | 2783 |
| $0.50 \leq z \leq 1$ | 142 | 23.90 | $-19.61 \pm 0.03$ | 0.330 | 0.7059 | 0.5737 | 819.0 | 3023 |
| $0.55 \leq z \leq 1.05$ | 125 | 24.05 | $-19.66 \pm 0.03$ | 0.316 | 0.7505 | 0.5866 | 837.4 | 3214 |
| $0.60 \leq z \leq 1.10$ | 105 | 24.19 | $-19.68 \pm 0.03$ | 0.306 | 0.7889 | 0.5971 | 852.4 | 3379 |
| $0.65 \leq z \leq 1.15$ | 86 | 24.36 | $-19.72 \pm 0.04$ | 0.303 | 0.8425 | 0.6108 | 872.0 | 3609 |
| $0.70 \leq z \leq 1.20$ | 73 | 24.49 | $-19.76 \pm 0.04$ | 0.343 | 0.8860 | 0.6212 | 886.8 | 3795 |
| $0.75 \leq z \leq 1.25$ | 67 | 24.58 | $-19.78 \pm 0.05$ | 0.362 | 0.9156 | 0.6279 | 896.4 | 3922 |
| $0.80 \leq z \leq 1.30$ | 60 | 24.68 | $-19.79 \pm 0.05$ | 0.369 | 0.9494 | 0.6353 | 906.9 | 4067 |
| $0.85 \leq z \leq 1.35$ | 47 | 24.84 | $-19.85 \pm 0.06$ | 0.379 | 1.0161 | 0.6489 | 926.4 | 4351 |
| $z \geq 0.85$ | 51 | 24.95 | $-19.84 \pm 0.06$ | 0.378 | 1.0484 | 0.6552 | 935.4 | 4490 |
| $z \geq 0.9$ | 43 | 25.01 | $-19.88 \pm 0.06$ | 0.380 | 1.0817 | 0.6614 | 944.2 | 4633 |
| $z \geq 0.95$ | 34 | 25.13 | $-19.89 \pm 0.07$ | 0.407 | 1.1228 | 0.6687 | 954.6 | 4809 |

**Table 3a**

| Name | $cz$ | $D$ |
|---|---|---|
| 1996h | 185871 | 2407 |
| 1996i | 170882 | 2360 |
| 1996j | 89938 | 1343 |
| 1996k | 113921 | 1766 |
| 1996u | 128911 | 1659 |
| 1995ao | 71950 | 1122 |
| 1995ap | 89938 | 1023 |
| 1996t | 71950 | 813 |
| 1997ce | 131909 | 1838 |
| 1997cj | 149896 | 2137 |
| 1997ck | 290799 | 4058 |
| 1995k | 143601 | 1695 |
| 1997ap | 248828 | 3497 |
| 1997am | 124714 | 1541 |
| 1997aj | 174179 | 2018 |
| 1997ai | 134907 | 1956 |
| 1997af | 173580 | 2606 |
| 1997ac | 95934 | 1221 |
| 1997r | 196964 | 3053 |
| 1997p | 141502 | 2178 |
| 1997o | 112122 | 2693 |
| 1997h | 157691 | 2140 |
| 1997g | 228742 | 3768 |
| 1997f | 173880 | 2354 |
| 1996cn | 128911 | 2393 |
| 1996cm | 134907 | 2393 |
| 1996cl | 248228 | 4037 |
| 1996ck | 196664 | 2772 |
| 1996ci | 148397 | 1773 |
| 1996cg | 146898 | 2066 |

**Table 3b**

| Name | $cz$ | $D$ |
|---|---|---|
| 1996cf | 170882 | 2232 |
| 1995ba | 116319 | 1664 |
| 1995az | 134907 | 1631 |
| 1995ay | 143900 | 2072 |
| 1995ax | 184372 | 2035 |
| 1995aw | 119917 | 1335 |
| 1995at | 196364 | 1971 |
| 1995as | 149297 | 2972 |
| 1995ar | 139404 | 2545 |
| 1995aq | 135806 | 2313 |
| 1994g | 127412 | 1429 |
| 1999fw | 83342 | 1156 |
| 1999fn | 143001 | 1698 |
| 1999fm | 284803 | 3107 |
| 1999fk | 316881 | 3939 |
| 1999fj | 244631 | 3264 |
| 1999ff | 136406 | 2324 |
| 2002ad | 154093 | 2011 |
| 2002ab | 126812 | 1660 |
| 2002aa | 283604 | 3808 |
| 2002x | 257522 | 4451 |
| 2002w | 309086 | 3279 |
| 2001kd | 280606 | 4900 |
| 2001jp | 158290 | 1796 |
| 2001jn | 193366 | 4717 |
| 2001jm | 293197 | 3476 |
| 2001jh | 265316 | 3283 |
| 2001jf | 244331 | 6230 |
| 2001iy | 170282 | 1933 |
| 2001ix | 213152 | 2700 |

**Table 3c**

| Name | $cz$ | $D$ |
|---|---|---|
| 2001iw | 101809 | 1347 |
| 2001iv | 118868 | 1577 |
| 2001hy | 243431 | 5357 |
| 2001hx | 239534 | 4832 |
| 2001hu | 264417 | 4938 |
| 2001hs | 249727 | 3307 |
| 2001fs | 262019 | 5706 |
| 2001fo | 231440 | 2492 |
| 2000fr | 162787 | 1927 |
| 1998bi | 224844 | 2808 |
| 1998be | 191867 | 2865 |
| 1998ba | 128911 | 1933 |
| 1998ay | 191867 | 2721 |
| 1998ax | 148997 | 2156 |
| 1998aw | 131909 | 2342 |
| 1998as | 106426 | 1840 |
| 1997ez | 233838 | 3451 |
| 1997eq | 161888 | 2089 |
| 1997ek | 257822 | 3752 |
| 04Eag | 305788 | 4635 |
| 04Gre | 341764 | 3617 |
| 04Man | 256023 | 3900 |
| 04Mcg | 410716 | 6080 |
| 04Omb | 292298 | 4507 |
| 04Pat | 290799 | 4973 |
| 04Rak | 221846 | 2706 |
| 04Sas | 416712 | 6355 |
| 04Yow | 137905 | 2945 |
| 05Fer | 305788 | 4214 |
| 05Gab | 335768 | 4620 |

**Table 3d**

| Name | $cz$ | $D$ |
|---|---|---|
| 05Lan | 368745 | 7739 |
| 05Red | 356753 | 6863 |
| 05Spo | 251526 | 3163 |
| 05Str | 302791 | 4862 |
| 05Zwi | 156192 | 2011 |
| 2002dc | 142401 | 2119 |
| 2002dd | 284803 | 3955 |
| 2002fw | 389730 | 6024 |
| 2002hp | 391229 | 5139 |
| 2002hr | 157691 | 3727 |
| 2002kd | 220347 | 3053 |
| 2002ki | 341764 | 5495 |
| 2003az | 379237 | 6265 |
| 2003dy | 401722 | 6339 |
| 2003eq | 251826 | 3492 |
| 03D4au | 140303 | 3494 |
| 04D4bk | 251826 | 3400 |
| 04D3nr | 287801 | 3619 |
| 04D3lu | 246370 | 3520 |
| 04D3ki | 278807 | 4632 |
| 04D3gt | 135206 | 2361 |
| 04D3do | 182873 | 2537 |
| 04D3cp | 248828 | 3271 |
| 04D2gp | 211953 | 3407 |
| 04D2fp | 124414 | 1606 |
| 04D1ag | 166984 | 1871 |
| 03D4fd | 237136 | 3308 |
| 03D4cz | 208356 | 3178 |
| 03D4at | 189769 | 2772 |
| 03D3bh | 74528 | 867 |

**Table 3e**

| Name | $cz$ | $D$ |
|---|---|---|
| 03D3af | 159490 | 2582 |
| 03D1fc | 99231 | 1151 |
| 03D1bp | 103728 | 1632 |
| 04D4dw | 288101 | 3690 |
| 04D4an | 183773 | 3391 |
| 04D3nh | 101989 | 1376 |
| 04D3lp | 294696 | 4635 |
| 04D3is | 212853 | 3659 |
| 04D3fq | 218849 | 3297 |
| 04D3df | 140903 | 2699 |
| 04D3co | 185871 | 2877 |
| 04D2gc | 156192 | 2343 |
| 04D2cf | 110623 | 1501 |
| 03D4gl | 171182 | 2163 |
| 03D4dy | 181075 | 2182 |
| 03D4cy | 277938 | 4192 |
| 03D4ag | 85441 | 875 |
| 03D3ba | 87300 | 1369 |
| 03D1gt | 164286 | 3852 |
| 03D1ew | 260220 | 3463 |
| 03D1ax | 148697 | 1924 |
| 04D4dm | 243132 | 3672 |
| 04D3oe | 226643 | 3122 |
| 04D3nc | 244930 | 3371 |
| 04D3ks | 225444 | 2750 |
| 04D3hn | 165366 | 2505 |
| 04D3fk | 107266 | 1691 |
| 04D3dd | 302791 | 5164 |
| 04D2ja | 222146 | 3203 |
| 04D2gb | 128911 | 1855 |

**Table 3f**

| Name | $cz$ | $D$ |
|---|---|---|
| 04D1ak | 157691 | 2840 |
| 03D4gg | 177477 | 2316 |
| 03D4di | 271312 | 3191 |
| 03D4cx | 284503 | 3552 |
| 03D3cd | 138114 | 1570 |
| 03D3ay | 111193 | 1385 |
| 03D1fq | 239834 | 4047 |
| 03D1co | 203559 | 3375 |
| 03D1aw | 174389 | 2633 |
| 04D4bq | 164886 | 2342 |
| 04D3ny | 242832 | 3390 |
| 04D3ml | 284803 | 3671 |
| 04D3kr | 101120 | 1256 |
| 04D3gx | 272811 | 4219 |
| 04D3ez | 78845 | 1147 |
| 04D3cy | 192767 | 2857 |
| 04D2iu | 207157 | 3719 |
| 04D2fs | 107026 | 1588 |
| 04D1aj | 216150 | 2856 |
| 03D4gf | 174179 | 2266 |
| 03D4dh | 187910 | 2236 |
| 03D4cn | 245230 | 4356 |
| 03D3aw | 134607 | 1577 |
| 03D1fl | 206257 | 2469 |
| 03D1cm | 260820 | 3673 |
| 03D1au | 151185 | 1933 |
| b010 | 177177 | 2318 |
| b013 | 127712 | 1734 |
| b016 | 98632 | 1707 |
| d033 | 159190 | 2197 |

**Table 3g**

| Name | $cz$ | $D$ |
|---|---|---|
| d058 | 174779 | 2630 |
| d084 | 155592 | 2877 |
| d085 | 120217 | 1580 |
| d087 | 101930 | 1212 |
| d089 | 130710 | 1550 |
| d093 | 108825 | 1175 |
| d097 | 130710 | 1550 |
| d117 | 92636 | 1605 |
| d149 | 102529 | 1413 |
| e029 | 99531 | 1722 |
| e108 | 140603 | 1550 |
| e132 | 71650 | 1186 |
| e136 | 105527 | 1994 |
| e138 | 183473 | 3463 |
| e140 | 189169 | 2228 |
| e147 | 193366 | 2189 |
| e148 | 128611 | 1700 |
| e149 | 148997 | 1855 |
| f011 | 161588 | 2264 |
| f041 | 168184 | 1968 |
| f076 | 122915 | 1472 |
| f096 | 123515 | 2205 |
| f216 | 179576 | 2874 |

**Table 3h**

| Name | $cz$ | $D$ |
|---|---|---|
| f231 | 185572 | 2335 |
| f235 | 126512 | 1525 |
| f244 | 161888 | 2276 |
| f308 | 120217 | 2242 |
| g005 | 65355 | 986 |
| g050 | 189769 | 1959 |
| g052 | 114821 | 1474 |
| g055 | 90537 | 2830 |
| g097 | 101930 | 1481 |
| g120 | 152894 | 1720 |
| g133 | 126213 | 2340 |
| g142 | 119617 | 2874 |
| g160 | 147798 | 1884 |
| g240 | 205957 | 2143 |
| h283 | 150496 | 2587 |
| h300 | 205957 | 2307 |
| h319 | 148397 | 1858 |
| h323 | 180775 | 2411 |
| h342 | 126213 | 1517 |
| h359 | 104328 | 1831 |
| h363 | 63856 | 1445 |
| h364 | 103129 | 1084 |
| k396 | 81244 | 1242 |

**Table 3i**

| Name | $cz$ | $D$ |
|---|---|---|
| k411 | 169083 | 1744 |
| k425 | 82143 | 1309 |
| k430 | 174479 | 3034 |
| k441 | 203859 | 2648 |
| k448 | 120217 | 2655 |
| k485 | 124714 | 3861 |
| m027 | 85741 | 1801 |
| m062 | 94135 | 1297 |
| m138 | 174179 | 2171 |
| m158 | 138804 | 2142 |
| m193 | 102229 | 1058 |
| m226 | 201161 | 2520 |
| n256 | 189169 | 2256 |
| n258 | 156492 | 2298 |
| n263 | 110324 | 1270 |
| n278 | 92636 | 1219 |
| n285 | 158290 | 2258 |
| n326 | 80344 | 1448 |
| p454 | 208356 | 2976 |
| p455 | 85141 | 1113 |
| p524 | 152295 | 1849 |
| p528 | 234138 | 3141 |
| p534 | 183773 | 2275 |